\def\rfr#1{eq. (\ref{#1})}

\def\virg#1{``#1''}

\def\bb#1#2#3{\bibitem[\protect\citeauthoryear{#1}{#2}]{#3}}
\def\eqi{\begin{equation}}
\def\eqf{\end{equation}}
\def\eqia{\begin{eqnarray}}
\def\eqfa{\end{eqnarray}}
\def\Om{\mathit{\Omega}}
\def\rp#1#2{{#1\over#2}} \def\lb#1{\label{#1}}
\def\bb#1#2#3{\bibitem[\protect\citeauthoryear{#1}{#2}]{#3}}

\def\bds#1{\boldsymbol{#1}}

\documentclass{elsart}

\usepackage{slashbox}
\usepackage{url}
\usepackage{amsmath}
\usepackage{latexsym,wasysym}

\usepackage{natbib}

\usepackage{epsfig}

\usepackage{amssymb}

\RequirePackage{color}

\begin{document}

\begin{frontmatter}



\title{Novel considerations about the error budget of the LAGEOS-based tests of frame-dragging with GRACE geopotential models}


\author{Lorenzo Iorio\thanksref{footnote2}}
\address{Ministero dell'Istruzione, dell'Universit\`{a} e della Ricerca (M.I.U.R.)-Istruzione\\
International Institute for Theoretical Physics and
Advanced Mathematics Einstein-Galilei\\ Fellow of the Royal Astronomical Society (F.R.A.S.)}
\corauth[cor]{Corresponding author}
\thanks[footnote2]{Address for correspondence:  Viale Unit\`{a} di Italia 68, 70125, Bari (BA), Italy}
\ead{lorenzo.iorio@libero.it}
\ead[url]{http://digilander.libero.it/lorri/homepage$\_$of$\_$lorenzo$\_$iorio.htm}

\author{Matteo Luca Ruggiero\thanksref{footnote3}}
\address{Dipartimento di Fisica, Politecnico di Torino, and INFN-Sezione
di Torino}
\thanks[footnote3]{Address for correspondence: Corso Duca Degli Abruzzi 24, 10129 Torino, Italy}

\author{Christian Corda\thanksref{footnote4}}
\address{International Institute for Theoretical Physics and
Advanced Mathematics Einstein-Galilei}
\thanks[footnote4]{Address for correspondence:  Via Santa Gonda 14, 59100, Prato, Italy}



\begin{abstract}
A realistic assessment of the uncertainties in the even zonals of a given geopotential model must be made by directly comparing its coefficients with those of a wholly independent solution of superior formal accuracy. Otherwise, a favorable selective bias is introduced in the evaluation of the total error budget of the LAGEOS-based Lense-Thirring tests yielding \textcolor{black}{likely} too optimistic
figures for it. By applying a novel approach \textcolor{black}{which} recently appeared in \textcolor{black}{the} literature,  the second ($\ell=4$) and the third ($\ell=6$) even zonals turn out to be uncertain at a $2-3\times 10^{-11}\ (\ell=4)$ and $3-4\times 10^{-11}\ (\ell=6)$ level, respectively, yielding a total gravitational error of about $27-28\%$, with an upper bound of $37-39\%$. The results by Ries \textit{et al.} themselves
yield an upper bound for it of about $33\%$.
The low-degree even zonals are not exclusively determined from the GRACE Satellite-to-Satellite Tracking (SST) range since they affect it with long-period, secular-like signatures over orbital arcs longer than one orbital period: GRACE SST is not accurately sensitive to such signals. Conversely, general relativity affects it with short-period effects as well. Thus, the issue of the a-priori \virg{imprinting} of general relativity itself in the GRACE-based models used so far remains open.
\end{abstract}


\end{frontmatter}

\parindent=0.5 cm


\section{Introduction}
According to the weak-field and slow-motion approximation of general relativity \citep{Rin}, matter-energy currents concur to generate  the overall gravitational field with an own peculiar contribution usually dubbed \virg{gravitomagnetic} \citep{Thorne88,Rin,Mash07} because of its formal resemblance to the magnetic field induced by electric currents in the framework of the linear Maxwellian electromagnetic theory. In the case of a slowly rotating body with proper angular momentum $\bds L$ like, e.g., our planet, its gravitomagnetic field \citep{Thorne86,Thorne88,Mash01}, proportional to $\bds L$, affects the motion of  a test particle orbiting it with a non-central, Lorentz-like force which causes small secular precessions of the longitude of the ascending node\footnote{$\Om$ is an angle, in the reference plane $\{x,y\}$,  between a given reference direction $x$ in it and the line of the nodes, which is the intersection of the satellite's orbital plane with the reference plane $\{x,y\}$ \citep{Roy}.} $\Om$
and the argument of pericenter\footnote{$\omega$ is an angle, in the orbital plane,  between the line of the nodes  and the point of closest approach to the central body, generally denoted as pericenter \citep{Roy}.} $\omega$ of its orbit: they are usually known as the Lense-Thirring effect \citep{LT}.

\citet{cugu1,cugu2}, for the first time, put forth the idea of measuring these effects in the terrestrial gravitational field by using  the non-dedicated passive geodetic satellite LAGEOS \citep{lageos},  launched in 1976 and tracked with the Satellite Laser Ranging (SLR) technique \citep{slr}, along with other SLR targets. The first practical attempts to implement such an idea date back to the mid 90s \citep{Ciu96a,Ciu97a,Ciu97,Ciu98,Ries03a}. In them,  the data of the nodes of LAGEOS and LAGEOS II \citep{lageos2}, launched in 1992, and the perigee of LAGEOS II were suitably dealt with according to a strategy proposed by \citet{Ciu96}. Actually, the orbits of the LAGEOS satellites are not only affected by the gravitomagnetic field of the Earth but also by a host of competing classical forces of gravitational \citep{kaula} and non-gravitational \citep{Milani} origin, so that a realistic assessment of the total error budget in such tests is not a trivial task. In the most recent attempts \citep{Ciu04,Ciu06,Riesetal}, summarized by their authors in \citep{Ciu07,Ciu09,Ciu010a,Ciu011b}, a linear combination of the nodes $\Om$ of both LAGEOS and LAGEOS II was adopted\footnote{See also \citet[Sec. 6]{Iorio07rebut}.} \citep{Pav02,Ries03a,Ries03b,Iorio04,Iorio06}. Such a combination was specifically designed to remove the biasing impact of the first even ($\ell=2,4,6,\ldots$) zonal ($m=0$) harmonic coefficient $J_2$ of the multipolar expansion of degree $\ell$ and order $m$ of the Newtonian part of the terrestrial gravitational potential \citep{kaula,Heisk}. However, the other even zonals of higher degree $J_4, J_6, J_8,\ldots$ do have an impact on the combination adopted depending on the level of mismodeling in them. Generally speaking, the even zonal multipoles of the geopotential induce biasing \textcolor{black}{of} secular precessions of the satellites' nodes which, for the lowest degrees, are nominally much larger than the Lense-\textcolor{black}{Thirring} ones. The normalized Stokes coefficients\footnote{The even zonals are defined as $J_{\ell}=-\sqrt{2\ell +1}\ \overline{C}_{\ell,0},\ \ell=2,4,6,\ldots$ } $\overline{C}_{\ell,m},\overline{S}_{\ell,m}$ of the geopotential are simultaneously estimated as solve-for parameters of global solutions \citep{icgem} in which huge amounts of data from dedicated \textcolor{black}{spacecrafts} are processed; latest improvements brought in by the ongoing Gravity Recovery and Climate Experiment (GRACE)  mission \citep{grace1,grace2} allowed \textcolor{black}{reduction of} the magnitude of the biasing \textcolor{black}{of the} even zonals signature  with respect to the earlier tests \citep{Ciu97,Ciu98,Ries03a}, when less accurate global gravity field models \citep{JGM3,EGM96} based on extensive data records of SLR satellites, \textcolor{black}{such as} LAGEOS and LAGEOS II themselves, were used.

In early 2012 the rocket VEGA launched LARES, a third passive geodetic satellite of LAGEOS-type\footnote{While the altitudes of LAGEOS and LAGEOS II are of the order of $5,800$ km, LARES orbit\textcolor{black}{s} at  $1,450$ km.},  which, according to the intentions of its proponents \citep{Ciu010},  should push the accuracy of the Lense-Thirring tests to $\approx 1\%$, provided that its data will suitably be combined with those from the already existing LAGEOS and LAGEOS II. In 1986, \citet{Ciu86} proposed to launch a new passive SLR target, LAGEOS X,  at the same altitude of LAGEOS, but with an orbital inclination $I$ to the Earth's equator differing  by 180 deg from that of LAGEOS. This choice was motivated by the fact that, in principle, all the secular precessions due to the even zonals would exactly cancel out in the sum of the nodes of the two satellites, contrary to the Lense-Thirring precessions which, instead, would sum up. The possibility of somewhat relaxing the original requirements \citep{Ciu86} on the orbital parameters of a third LAGEOS satellite, to be used in combination with LAGEOS and LAGEOS II to measure the Lense-Thirring effect,  was envisaged by \citep{IorioNA05b} in view of the expected improvements in determining the geopotential multipoles from GRACE.

In all the LAGEOS-based tests performed so far, the existence of the Lense-Thirring signal was always indirectly inferred  by constructing time series of post-fit computed node \virg{residuals} for LAGEOS and LAGEOS II, and by fitting the resulting time series with a straight line and other harmonic signals. A total accuracy of $\approx 10\%$ has always been claimed \citep{Ciu07,Ciu09,Ciu010a,Ciu011b}. However, in all such tests,  the gravitomagnetic field of the Earth was never modeled nor explicitly solved-for together with other parameters routinely estimated in the data reduction procedure. Moreover, general relativity was never solved-for in all the GRACE-based global Earth gravity field solutions which are used as background models for the geopotential  for LAGEOS and LAGEOS II. Thus, several aspects of such tests have been considered unsatisfactory by us: e.g., the total error budget may be up to $2-3$ times larger than proposed by \citet{Ciu07,Ciu09,Ciu010a,Ciu011b} for a number of reasons. Moreover, also the overall accuracy obtainable with LARES may be worse than the claimed $\approx 1\%$ level.

\citet{Iorio011}, among other aspects of gravitomagnetism in the solar system, summarized the main critical points of the tests performed with the LAGEOS \textcolor{black}{spacecrafts} so far. \citet{Ciu011} extensively criticized this part of \citet{Iorio011}.
According to \citet[p. 341]{Ciu011}, none of the claims by \citet{Iorio011} could be reproduced by any of their independent analyses. Actually,
the paper by \citet{Ciu011} does not contain any  real progress from the scientific point of view since  they  substantially limited themselves to repeat again
some of their arguments already exposed elsewhere without adding  new quantitative elements backing them, and leaving many of the remarks by \citet{Iorio011} unaddressed. Furthermore, some of the claims by \citet{Ciu011} do not find  support in the existing scientific literature.

In this paper, we offer \textcolor{black}{novel and quantitative}  arguments supporting our positions about certain important points concerning the interplay among the Newtonian geopotential and the general relativistic frame-dragging.
\section{On the accuracy of the GRACE even zonal harmonics}\lb{mache}
\subsubsection{The LAGEOS-LAGEOS II tests}\lb{quinta}
As a general remark concerning a realistic evaluation of the error budget in satellite-based tests of the Lense-Thirring effect with LAGEOS-type satellites, \citet[p. 26]{Ruffini03} recall that Shapiro in many discussions puts a factor from experience that in this general style of modeling, one needs  to take the error given by the computer error model and put in a real-life correction factor of 2 to 3.
 Incidentally, a straightforward application of such words to the
 claimed $\approx 10\%$ total error by \citet{Ciu07,Ciu09,Ciu010a,Ciu011b} would yield figures for it as large as just those proposed  by \citet{Iorio011}.

The existing scientific literature does not support the remarks by \citet[p. 343]{Ciu011}, which we consider
ad-hoc
attempts to pick up just those models which better satisfy the needs of \citet{Ciu011} in view of the level of accuracy desired for their tests.
Indeed, the idea of estimating field errors by taking the difference among geopotential coefficients (without particular regard to independence or quality of the solutions from which they are retrieved) is an old one. See, e.g., \citet{Martin70,Lerch91,Lerch94}, Fig. 9-11 of \citet[pp. 90-92]{cinciaolin}, where the differences among the estimated geoid heights\footnote{They are  expressed in terms of the geopotential multipolar coefficients \citep{Heisk}.} from simulated GRACE data records are compared to those from EGM96 \citep{EGM96} along with their $\sigma$ and $3\sigma$ confidence bounds,
the figures displayed in \citet{eigencg01c} \textcolor{black}{illustrating} the degree amplitudes of the differences between  EIGEN-CG01C \citep{eigencg01c} and EGM96 \citep{EGM96}. \textcolor{black}{Also of relevance, are}
the comparison by degree differences of the geoid heights of the EGM96 \citep{EGM96}, EGM2008 \citep{egm2008} and EIGEN-5C \citep{GFZ2} models by \citet{Yil}
and the recent discussion in \citep{Wagner011}.
\citet{icgem} itself, at the page \virg{Evaluation of Models}, plots the amplitudes of the difference per degree of several models, many of them dating back even to the pre-CHAMP/GRACE/GOCE era, to the recent -- and formally much more accurate -- combined solution\footnote{Such a particular model would not be suited for the LAGEOS-based tests of the Lense-Thirring effect since it includes data of LAGEOS itself.} EIGEN-6C \citep{GFZ3}.
A close examination of the content of \citet{Wagner011} is instructive.
\citet{Wagner011}, who feel the need of comparing not only solutions releasing the mere statistical, formal errors for the geopotential coefficients, but also older models yielding calibrated errors, explicitly require that the benchmark model must be formally far more accurate than the one to be tested.  Furthermore, \citet{Wagner011} point out that their method could be applied well even to solutions not displaying formal errors.
 More precisely, the aim of \citet{Wagner011}  is to test the accuracy of some recent models' formal error estimates by comparing the field coefficients directly with the same coefficients from a wholly independent and formally superior model. To validate their method, \citet{Wagner011}  also apply it to fairly recent historic models which underwent extensive calibration. However, even without formal variances for a test harmonic field, \citet{Wagner011} could approximate them by comparing only the coefficients themselves, those of the test with a clearly superior and independent reference model. Still, to gain further confidence, \citet{Wagner011} include comparisons with two older models from combined multiple-satellite conventional tracking and surface anomalies from ground survey and altimetry: JGM3, $70\times 70$, \citep{JGM3} and EGM96, $360\times 360$, \citep{EGM96}.
 \citet{Wagner78} apply such an idea of coefficient error calibration without requiring
(as in \citet{Wagner011}) that the reference model be both independent and
clearly superior to the calibrated one.  As another example, \citet[p. 17]{Milani} evaluated the uncertainty in  $J_2$ just by  taking the difference between the estimated values for it from two global gravity field models, i.e. GEM-L2 \citep{GEML2} and GEM 9 \citep{GEM9}: while the uncertainty of the GEM 9 value was $\sigma_{J_2}=1\times 10^{-9}$, GEM-L2 yielded $\sigma_{J_2}=4\times 10^{-10}$. The same approach  was followed by \citet{compa} with the model GEM-L2 \citep{GEML2}, as \textcolor{black}{correctly} recognized by
\citet[p. 1713]{Ciu96} himself. Incidentally, \citet[p. 13]{Milani} adopted the same approach to evaluate the uncertainty in the Earth's gravitational parameter $GM$ as well.
\citet[p. 1713]{Ciu96} and \citet[p. 2712]{Ciu97} took the differences among the coefficients of the models JGM3 \citep{JGM3} and GEM-T3S \citep{Lerch94}, the latter being older and formally about one order of magnitude less accurate than JGM3.

Even the approach followed by \citet{Iorio011} may turn out to be intrinsically too optimistic and somewhat favorably biased. Indeed, \citet{Wagner011} point out that the calibration of the errors in a given test model should be made by using reference solutions obtained independently: more specifically, a GRACE-based solution should be compared with, say, a GOCE-based solution, as done by \citet{Wagner011}. Even in such a case, care should be taken to avoid that the reference model adopted was not used as a-priori background model in producing the models to be tested \citep{Wagner011}. Instead, all the models compared in  \citet{Iorio011} were obtained from GRACE itself.
In the following, we apply the method of \citet{Wagner011} by choosing the GRACE-based models GGM03S \citep{ggm03s} and EIGEN-GRACE02S \citep{eigengrace02s} as test models, while we take the wholly independent\footnote{Indeed, the background gravity models adopted for AIUB-CHAMP03S were JGM3 \citep{JGM3} and EGM96 \citep{EGM96}.  See the discussion in \citet{Wagner011} about the risk of precluding an unbiased calibration employing an external standard model.} CHAMP-based solution AIUB-CHAMP03S \citep{aiubchamp03s} as formally superior reference model.
In particular, we look at the even zonals of degree $\ell=4,6$ which are the most important ones in determining the error of gravitational origin in the performed LAGEOS-based tests of the Lense-Thirring effect. As requested by \citet{Wagner011}, the sigmas of AIUB-CHAMP03S are smaller than those of the models to be tested by about 1 order of magnitude in the degree range chosen $(\ell=4,6)$. In Table \ref{paragone} we display the corresponding error factors $f$ computed according to Eq. (A11) of \citet{Wagner011}
\eqi f_{\rm test, \ell}=\rp{\sqrt{\left(\overline{C}_{\ell,0}^{\rm test}-\overline{C}_{\ell,0}^{\rm ref}\right)^2-\left(\sigma^{\rm ref}_{\overline{C}_{\ell,0}}\right)^2}}{\sigma^{\rm test}_{\overline{C}_{\ell,0}}};\lb{tyu}\eqf the sigmas $\sigma^{\rm test}_{\overline{C}_{\ell,0}}$ of the test models have to be rescaled by such $f_{\rm test,\ell}$. In our case, \rfr{tyu} worked properly in the sense that it did not yield imaginary results.
\begin{table*}[ht!]
\caption{Application of the method by \citet{Wagner011} to the CHAMP-based model AIUB-CHAMP03S \citep{aiubchamp03s}, assumed as formally superior reference model, and the GRACE-based models GGM03S \citep{ggm03s} and EIGEN-GRACE02S \citep{eigengrace02s} as test models. The even zonal coefficients examined are $\overline{C}_{4,0}$ and $\overline{C}_{6,0}$. The formal errors of AIUB-CHAMP03S for them are $\sigma_{\overline{C}_{4,0}}=8\times 10^{-13}$ and $\sigma_{\overline{C}_{6,0}}=9\times 10^{-13}$, respectively. The (calibrated) errors of GGM03S are $\sigma_{\overline{C}_{4,0}}=4.2\times 10^{-12}$ and $\sigma_{\overline{C}_{6,0}}=2.2\times 10^{-12}$, respectively. The (calibrated) errors of EIGEN-GRACE02S are $\sigma_{\overline{C}_{4,0}}=3.9\times 10^{-12}$ and $\sigma_{\overline{C}_{6,0}}=2.0\times 10^{-12}$, respectively.
 Eq. (A11) of \citet{Wagner011} was used to compute the scaling error factors $f_{\rm GGM03S}$ and $f_{\rm EIGEN-GRACE02S}$ to be applied to the sigmas of GGM03S and EIGEN-GRACE02S.
}\label{paragone}
\centering
\bigskip
\begin{tabular}{lll}
\hline\noalign{\smallskip}
Even zonal coefficient & $f_{\rm GGM03S}$ & $f_{\rm EIGEN-GRACE02S}$  \\
\noalign{\smallskip}\hline\noalign{\smallskip}
$\overline{C}_{4,0}$ & $14.2$ & $16.2$ \\
$\overline{C}_{6,0}$ & $48.6$ & $67.2$ \\
\noalign{\smallskip}\hline\noalign{\smallskip}
\end{tabular}
\end{table*}
In Table \ref{paragone2} we repeat the same calculation by considering AIUB-CHAMP03S \citep{aiubchamp03s} as test model, to be calibrated by the formally superior and wholly independent GRACE-based solution ITG-Grace02s \citep{itggrace02s}.
\begin{table*}[ht!]
\caption{Application of the method by \citet{Wagner011} to the GRACE-based model ITG-Grace02s \citep{itggrace02s}, assumed as formally superior reference model, and the CHAMP-based model AIUB-CHAMP03S \citep{aiubchamp03s}  as test solution. The even zonal coefficients examined are $\overline{C}_{4,0}$ and $\overline{C}_{6,0}$. The formal errors of ITG-Grace02s are $\sigma_{\overline{C}_{4,0}}=8.6\times 10^{-14}$ and $\sigma_{\overline{C}_{6,0}}=4.5\times 10^{-14}$, respectively. The formal errors of AIUB-CHAMP03S for them are $\sigma_{\overline{C}_{4,0}}=8\times 10^{-13}$ and $\sigma_{\overline{C}_{6,0}}=9\times 10^{-13}$, respectively.
 Eq. (A11) of \citet{Wagner011} was used to compute the scaling error factors $f_{\rm AIUB-CHAMP03S}$  to be applied to the sigmas of AIUB-CHAMP03S.
}\label{paragone2}
\centering
\bigskip
\begin{tabular}{ll}
\hline\noalign{\smallskip}
Even zonal coefficient & $f_{\rm AIUB-CHAMP03S}$  \\
\noalign{\smallskip}\hline\noalign{\smallskip}
$\overline{C}_{4,0}$ & $62.2$ \\
$\overline{C}_{6,0}$ & $112.2$  \\
\noalign{\smallskip}\hline\noalign{\smallskip}
\end{tabular}
\end{table*}
From the results of Table \ref{paragone} and Table \ref{paragone2} it turns out that the second and the third even zonals are, actually, uncertain at a level
\eqi
\begin{array}{lll}
\delta {\overline{C}_{4,0}} &  = & f_{\rm test, 4}\times \sigma_{\overline{C}_{4,0}^{\rm test}}\approx  5-6\times 10^{-11}, \\ \\
\delta {\overline{C}_{6,0}} & = & f_{\rm test, 6}\times \sigma_{\overline{C}_{6,0}^{\rm test}}\approx  1\times 10^{-10}.
\end{array}\lb{inculo}
\eqf
It is remarkable how \rfr{inculo} comes from the application of the $f_{\ell}$ scaling factors of Table \ref{paragone} to the sigmas of both GGM03S and EIGEN-GRACE02S, and of the $f_{\ell}$ scaling factors of Table \ref{paragone2} to the sigmas of AIUB-CHAMP03S.
As expected, the uncertainties of \rfr{inculo} are generally less favorable than those obtained from the mutual comparisons of only GRACE-based models in \citet{Iorio011}.
Uncertainties as large as those in \rfr{inculo} correspond to a mismodeled competing signal from the zonals amounting to $66\%$ of the expected Lense-Thirring signal from a Root-Sum-Square (RSS) calculation.
However, \citet{Wagner011} offer also a different way to compute $f_{\rm test,\ell}$. Instead of using a single coefficient error scaling factor, they propose to average the individual error factors over all the $2\ell +1$ coefficients of degree $\ell$. According to Eq. (A13) of \citet{Wagner011}, one has
\eqi \overline{f}_{\rm test, \ell}=\left\{\left(\rp{1}{2\ell +1}\right)\sum_{m=0}^{2\ell+1}\left[\rp{\left(\overline{H}_{\ell,m}^{\rm test}-\overline{H}_{\ell,m}^{\rm ref}\right)^2-\left(\sigma^{\rm ref}_{\overline{H}_{\ell,m}}\right)^2}{\left(\sigma^{\rm test}_{\overline{H}_{\ell,m}}\right)^2}\right]\right\}^{\rp{1}{2}},\lb{A13}\eqf
where $\overline{H}_{\ell,m}$ denotes both $\overline{C}_{\ell,m}$  and $\overline{S}_{\ell,m}$ in the sense that the sum in \rfr{A13} includes all the geopotential coefficients of both kinds for a given degree $\ell$. It turns out that \rfr{A13} gives smaller uncertainties than \rfr{tyu} for the second and the third even zonals, especially as far as $\ell=6$ is concerned. Indeed, applying \rfr{A13} to EIGEN-GRACE02S \citep{eigengrace02s}, GGM03S \citep{ggm03s}, AIUB-CHAMP03S \citep{aiubchamp03s} and ITG-Grace02s \citep{itggrace02s} in the same roles
%
%
%
%
%
\eqi
\begin{array}{lll}
\delta {\overline{C}_{4,0}} & = & \overline{f}_{\rm test, 4}\times \sigma_{\overline{C}_{4,0}^{\rm test}}\approx 2-3\times 10^{-11}, \\ \\
\delta {\overline{C}_{6,0}} & = & \overline{f}_{\rm test, 6}\times \sigma_{\overline{C}_{6,0}^{\rm test}}\approx  3-4\times 10^{-11}.
\end{array}\lb{inculo2}
\eqf
The resulting total  mismodeled signal in the LAGEOS-LAGEOS II node combination is, thus, approximately $27-28\%$ (RSS)  of the Lense-Thirring expected signature, with an upper bound of about $37-39\%$ from a Sum of the Absolute Values (SAV) calculation.

The analysis by \citet{Riesetal} mentioned in (a) of \citet{Ciu011} was  already critically discussed in \citet[p.28-29]{CEJP2008}. Here we recall that
\citet{Riesetal} considered models \textcolor{black}{such as} EIGEN-GL04C \citep{GFZ1} and EIGEN-GL05C \citep{GFZ2} which include data from LAGEOS itself. We also remark  that some of the models adopted by \citet{Riesetal}, like GIF22a and JEM04G, are not publicly available.
Moreover, we are neither saying that the method followed by \citet{Riesetal} is, in principle, unsuitable nor that they made some technical mistakes in their RSS calculation:
it \textcolor{black}{seems}
tailored to yield too optimistic figures for the error of gravitational origin in the LAGEOS-based tests. Indeed, the table in \citet[p. 16]{Riesetal} yields an upper bound of $21\%$ for the error of gravitational origin (SAV calculation). Moreover,
from a
visual inspection of the scatter of the individual points in Figure 6 of \citet[p. 84]{Ciu09} and in Figure 5 of \citet[p. 9]{Ciu011b}, based on the work by \citet{Riesetal}, inferring from it an overall test uncertainty less than $15\%$ appears \textcolor{black}{optimistic} due to an a priori selection bias. Indeed, by halving the difference between the maximum ($\approx 1.38$) and the minimum ($\approx 0.72$) possible values reported in Figure 6 of \citet[p. 84]{Ciu09} and in Figure 5 of \citet[p. 9]{Ciu011b}, one gets an uncertainty as large as $\approx 33\%$. We did not make any a priori selection: even if we wanted to do so, it would be impossible since the points displayed in
Figure 6 of \citet[p. 84]{Ciu09} and in Figure 5 of \citet[p. 9]{Ciu011b} are not explicitly associated to any specific Earth's gravity model. Even by arbitrarily discarding the two less favorable points in Figure 6 of \citet[p. 84]{Ciu09} and in Figure 5 of \citet[p. 9]{Ciu011b}, the remaining largest possible value ($\approx 1.18$) and the smallest possible value ($\approx 0.74$) yield an uncertainty of about $22\%$.
Thus, we conclude that the results by \citet{Riesetal} themselves
are incompatible with a $10\%$ accuracy.

From a broader point of view, a straightforward and uncritical extension of standard approaches usually followed with success in satellite geodesy to the issue of performing genuine and unbiased tests of fundamental physics should be avoided. Indeed, if the goal of a satellite-based mission is, say,  making accurate remote sensing, then, the sole scope of using a certain background Earth gravity model in processing the satellite's data is minimizing its post-fit residuals in order to predict the satellite's path with the highest possible accuracy. In this respect, the way in which the parameters entering the Earth gravity model used have been obtained has no relevance: it works well, and that is all. On the contrary, this is not the case for an unbiased test of general relativity; otherwise, an a-priori favorable bias is introduced, driving the outcome of the tests just towards the expected (and desired) result. In this respect,
it is \textcolor{black}{not appropriate} to pick up, say, just those Earth gravity models which behave better than others in reducing the satellite post-fit residuals, or accurately select
those solutions yielding the smallest error budget in view of their published errors, calibrated or not. Moreover, there are \textcolor{black}{no}
sound reasons in principle, to consider only the GRACE-based models with respect to other global solutions obtained from other \textcolor{black}{spacecraft}, provided that they do not include LAGEOS and LAGEOS II themselves.
\subsubsection{The LARES test}\lb{sesta}
%
%

\citet[point (c), p. 343]{Ciu011} claims that -- based on their selected gravity models -- the  final accuracy of the LARES experiment will be at the level of a few percent ($2\sigma$ level) in 2017.
\textcolor{black}{W}e doubt this prediction for the following reasons. \textcolor{black}{For an independent analysis, which essentially supports our points, see \citet{Renzetti012}.}

Concerning the asserted certainties by \citet[point (c), p. 343]{Ciu011} about steady improvements in the GRACE-based Earth gravity models by the expected epoch of the LARES data analysis (2017), at the moment the current trend for the new GRACE-based global gravity field models \citep{icgem} points toward the production of solutions which, actually, could not be used in any LAGEOS-based tests of fundamental physics. Indeed, the latest GRACE-based models   EIGEN-6 \citep{GFZ3}, GOCO02S \citep{goco2}, EIGEN-5 \citep{GFZ2} \textcolor{black}{and} EIGEN-GL04 \citep{GFZ1}, include data from LAGEOS itself. It is just the case to recall that, actually, the aim of GOCE \citep{goce} is to improve the knowledge of the very short wavelength sector of the geopotential, corresponding to the multipoles of very high degree and orders\textcolor{black}{;} the longer wavelengths, corresponding to the relatively low-mid degree spherical harmonics which are of interest here, are left substantially unaffected by GOCE. The inclusion of data from LAGEOS itself and other SLR targets is aimed to improve just the zonals of low degrees (see also the discussion in Section \ref{settima}). Thus, it is not unlikely that, in principle, also LARES itself may be further included to produce new global gravity solutions which, of course, could not be employed as background models to test the Lense-Thirring effect with the LAGEOS satellites themselves.


Figure 1 displayed in \citet[p. 344]{Ciu011}, which is also shown in several other papers \citep{Ciu09,Ciu010,Ciu011b}, presentations, talks, etc., by the same authors, is considered to illustrate the error of gravitational origin in the LARES mission.
\textcolor{black}{I}t is based only on two global gravity field solutions. \textcolor{black}{C}ontrary to \citet{Iorio09a,Iorio09b}, no details at all were released by \citet{Ciu011} concerning the computational approach followed to obtain it.
%
%
%
%
%
Did \citet{Ciu011} numerically integrate the equations of motion of the
 LAGEOS satellites by simulating the orbit of LARES? If so, what are the approximations used in such a numerical calculation? What are its details? Or, instead, was an analytical approach followed? In this case, what theoretical scheme was adopted to compute the coefficients $\dot\Om_{.\ell}, \ell=2,4,6,\ldots$ of the secular node precessions induced by the even zonals? Knowing the details would be crucial to assess the claims by \citet{Ciu011}. Figure 1 of
 \citet[p. 3]{Jconfs}, equals to Fig. 8  of \citet[p. 87]{Ciu09} and to Fig. 6 of \citet[p. 13]{Ciu011b}, is accompanied by some more details. \citet{Jconfs} specify that only the effects of
the first 5 even zonal harmonics were considered, and stating-without any quantitative arguments-that including higher degree even
zonal harmonics, the results of Figure 1 would only change slightly. Incidentally, this fact proves that, actually, \citet{Ciu011} did not address at all the remark by \citet{Iorio011}.


The
fact that different authors,
%
%
%
%
%
%
 with different computational approaches, softwares and levels of truncation of
the even zonals, independently obtain so different and scattered values for the gravitational error in the LARES test which demonstrates that we are still far from having a reliable and unambiguous answer to this important issue. Indeed, opting for the computational scheme yielding just the best (and desired in advance) result would be another example of
selective bias.  It is unclear why one should a priori decide  that a calculation yielding an uncertainty larger than $1\%$ is unreliable, and accept the calculation giving just $\sim 1\%$. On the other hand, it is important to remark that, at present, there are no other independent reasons to judge our calculation unrealistic. Indeed, their outcome can be considered large just with respect to the Lense-Thirring effect itself, not to
the overall orbit which, indeed, would not be displaced by an unacceptably large amount.

\section{The issue of a conceivable imprint of the Lense-Thirring effect
in the even zonal harmonics}\lb{settima}
\citet[p. 344]{Ciu011} make
qualitative statements about GRACE and the issue of a possible \virg{imprint} of general relativity on the global Earth's gravity models produced from GRACE data: no quantitative analyses like, e.g., analytical calculation and/or numerical tests with real/simulated data are offered to support such claims. For details of our unaddressed points, see \citet{IorioCN,Iorio011,IorioSST}. In particular,  the preliminary numerical analysis in \citet{IorioSST} has shown, without limiting
to the gravitomagnetic field of the Earth, that general relativity, which has never been explicitly solved-for so far in any GRACE-based global solution, may actually have a non-negligible impact on the GRACE intersatellite  tracking as well. In general, \citet{Ciu011} seem to confuse the orbital effects that a given force induces on the LAGEOS satellites and which are pertinent to the relativity tests, with those affecting the GRACE intersatellite  dynamics. To this aim,  the statements by \citet[p. 344]{Ciu011} concerning the fact that low-degree even zonal harmonics would be almost exclusively determined by means of the GRACE Satellite-to-Satellite Tracking (SST)   \textcolor{black}{may be} incorrect; indeed, from a general point of view,
 the intersatellite signal has significant information on the medium to shorter wavelength \citep[p. 2]{eigengrace02s}, corresponding to medium-higher degrees in terms of spherical harmonics.
According to \citet[p. 2]{eigengrace02s}, notable improvements in determining the long-wavelength coefficients, with respect to the pre-CHAMP/GRACE era,  occurred from the high-low GPS-CHAMP orbit tracking, which is used for GRACE as well. Moreover, \citet{Wagner011} explicitly state that, actually, the GPS orbit data for both the GRACE satellites are used as observations in the GRACE models, playing a role mainly just at the lowest degrees since their accuracy is at the cm-level.
On the other hand, if  the low-degree even zonals were really so accurately determined from the GRACE intersatellite tracking only, there would be no need of including also data from LAGEOS itself just to improve the knowledge of the long wavelength terms over timescales long enough to completely average out their seasonal\footnote{\citet[p.8]{eigengrace02s} write that it could  partly explain why the accuracy in the GRACE-based models at the very long-wavelength scale has not yet reached the anticipated baseline accuracy.}  temporal variations, as the current trend of the latest global gravity field solutions \citep{icgem} clearly shows. More specifically, \citet[p. 344]{Ciu011} write that the inter-satellite range measurement on GRACE is so accurate that it is able to track the short-period variations associated with the even zonal harmonics. Actually, \textcolor{black}{it is shown in} Figure 4 of \citet{IorioSST} that the low-degree even zonals of interest for the LAGEOS-based tests do induce long-period, secular-like signatures in the GRACE SST range over typical arcs 1 d long, while the orbital period of the GRACE satellites is $1.56\ {\rm h}= 0.065\ {\rm d}$.
%
%
%
%
%
%
%
%
%
%
%
%
Moreover, \citet[p. 344]{Ciu011} claim that modeling general relativity or not would be irrelevant for the GRACE models since the relativistic signals would be secular and/or long-period, and that the measurement type, i.e. the GRACE SST tracking, is relatively insensitive to secular or
long-period signals (essentially any signal with period of one revolution or longer). Actually,
Figure 3 of \citet{IorioSST} clearly \textcolor{black}{illustrates}
that the 1PN gravitoelectric Schwarzschild-like SST range signal exhibit relevant short-period patterns as well. On the other hand, in view of Figure 4 of \citet{IorioSST},  the previously cited statement by \citet{Ciu011} concerning the relative inability of the GRACE SST tracking to measure secular or, in the aforementioned specified sense, long-period signals  just confirms indirectly that the low-degree even zonals are  not exclusively determined from the GRACE intersatellite tracking itself.
\section{Summary and conclusions}\lb{conclusione}
\citet{Ciu011}, in their  assessments of the total error budget, rated at $10\%$,
make
selective choices of just those Earth gravity models yielding the desired result. \textcolor{black}{S}ome of these models are not
publicly available, while some others included data from LAGEOS itself. Also from the point of view of the methods used to compute the overall gravitational error from their
selected models, \citet{Ciu011} apply approaches which
tend to selectively and systematically reduce this error.
The claims by \citet{Ciu011} about the choice of the models to be used and how to compare them are, actually, not supported in the existing scientific literature. Indeed, it is widely recognized that the most conservative and reliable approach to realistically evaluate the uncertainties in the geopotential coefficients consists of directly comparing their values obtained in different global Earth gravity models without limiting to just those of comparable accuracy, as incorrectly claimed by \citet{Ciu011}. On the contrary, it has recently been shown  that it is necessary to use reference models having statistical, formal errors much smaller than those of the solutions to be tested \citep{Wagner011}. Such a procedure must be applied not only to those solutions releasing just formal variances, but also to those having calibrated errors as well \citep{Wagner011}. Finally, in order to further reduce any favorable bias, as it may have occurred in \citet{Iorio011}, where only GRACE-based models were reciprocally compared, the confrontation must be made among models which have been obtained quite independently \citep{Wagner011}.
A consistent application of such recent methods \citep{Wagner011} to different, independently obtained Earth gravity models yields uncertainties in the second and the third even zonals as large as $\delta\overline{C}_{4,0}\approx 5-6\times 10^{-11}$ and $\delta\overline{C}_{6,0}\approx1\times 10^{-10}$, respectively. Other, more favorable evaluations are $\delta\overline{C}_{4,0}\approx 2-3\times 10^{-11}$ and $\delta\overline{C}_{6,0}\approx3-4\times 10^{-11}$, respectively\textcolor{black}{;} they are based on a variant of the method exposed in \citet{Wagner011} involving an average over all the coefficients of a given degree. It turns out that, even with such more favorable errors in the zonals, the total systematic uncertainty of gravitational origin in the LAGEOS-based tests is about $27-28\%$, with an upper bound of approximately $37-39\%$. The results by \citet{Riesetal} themselves, displayed in
Figure 6 of \citet[p. 84]{Ciu09} and in Figure 5 of \citet[p. 9]{Ciu011b}, yield an upper bound for it of about $33\%$ from the scatter of all their points, although \citet{Riesetal} either used some models including data from LAGEOS  itself or solutions not publicly available  in \citet{icgem}. Even by arbitrarily rejecting the two points exhibiting the largest discrepancies with respect to the desired outcome, the remaining points by \citet{Riesetal} yield an uncertainty of about $22\%$. Figure 6 in \citet[p. 84]{Ciu09} and Figure 5 in \citet[p. 9]{Ciu011b}, based on the work by \citet{Riesetal}, are unambiguous in this respect.
Thus, the claims of a total $10\%$ error in the LAGEOS-based tests are \textcolor{black}{optimistic}.

The current trend in producing global gravity fields by several independent international institutions points toward the generation of models including data from LAGEOS as well to improve just the even zonals of very low degrees\textcolor{black}{;} GOCE was designed to accurately determine the multipoles of very high degree and order. Thus, such new global solutions cannot be used as background geopotential models in any present and future test of fundamental physics using just the data of the LAGEOS satellites themselves as primary source.

The claims by \citet{Ciu011} concerning the $\approx 1\%$ level of the expected gravitational error in the LARES experiment should  be supported with explicit and quantitative details about the procedures used in their evaluations.
\textcolor{black}{It is unclear why one should a priori consider as acceptable a (still undisclosed) computational scheme yielding just the desired $1\%$, and automatically reject  as unreliable the calculations by other authors that provide more details and make provision for a larger uncertainty. }

In dealing with the issue of the possible, a-priori \virg{imprinting} of general relativity itself in the GRACE-based models, it seems that \citet{Ciu011}
confuse some dynamical orbital effects affecting the LAGEOS satellites with those actually occurring for the GRACE intersatellite tracking. Thus, \citet{Ciu011} 
conclude that the low-degree even zonal harmonics are almost exclusively determined from GRACE SST range since it is able to accurately measure just short-periods signals like those allegedly induced on it by the extremely long-wavelength components of the geopotential. Moreover, \citet{Ciu011} state that including general relativity or not in the GRACE models would be irrelevant because of its allegedly long-term effects, not sensed by the GRACE SST range. We showed that this is not so. Indeed, if, on the one hand, the low-degree even zonals affect the GRACE SST range with long-period, secular-like effects, on the other hand, general relativity, never explicitly solved-for in all the GRACE-based models produced so far, \textcolor{black}{also induces} also high-frequency signatures on the GRACE SST range. Thus, the issue of the general relativistic \virg{imprint} remains open.



\end{document}